# The allostery landscape: quantifying thermodynamic couplings in biomolecular systems.

*Michel A. Cuendet[‡§‡], Harel Weinstein[†§], Michael V. LeVine[†§*]*

[†]Department of Physiology and Biophysics and [§]HRH Prince Alwaleed Bin Talal Bin Abdulaziz Alsaud Institute for Computational Biomedicine, Weill Cornell Medical College of Cornell University, New York, New York 10065, United States

[‡]Swiss Institute of Bioinformatics, UNIL Sorge, 1015 Lausanne, Switzerland

[*]E-mail: mil2037@med.cornell.edu

Allostery plays a fundament role in most biological processes. However, little theory is available to describe it outside of two-state models. Here we use a statistical mechanical approach to show that the allosteric coupling between two collective variables is not a single number, but instead a two-dimensional thermodynamic coupling function that is directly related to the mutual information from information theory and the copula density function from probability theory. On this basis, we demonstrate how to quantify the contribution of specific energy terms to this thermodynamic coupling function, enabling a decomposition that reveals the mechanism of allostery. We illustrate the thermodynamic coupling function and its use by showing how allosteric coupling in the alanine dipeptide molecule contributes to the overall shape of the $\Phi/\Psi$ free energy surface, and by identifying the interactions that are necessary for this coupling.

# Introduction

Allostery plays a fundament role in most biological processes, and has been suggested to be present in nearly all proteins[1]. One of the best-studied allosteric phenomena is the activation of a receptor, which we will denote as R, by a ligand, denoted as L. The most common model for allostery in this system is the allosteric two-state model (ATSM)[2,3]. We can construct a thermodynamic cycle for the process of ligand-induced activation of the receptor:

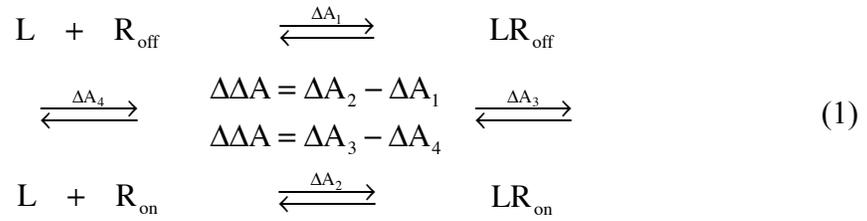

$$(1)$$

We refer to this as an "allosteric cycle". To describe the allostery in this system, the allosteric efficacy, $\alpha$, can be calculated from the cycle as

$$\alpha = \frac{K_{bound}}{K_{unbound}} \ , \qquad (2)$$

where (assuming the volume is constant), the equilibrium constants are a function of the differences in Helmholtz free energy, A, for the two states:

$$K = e^{-\beta \Delta A} \qquad (3)$$

Thus,

$$\alpha = \frac{e^{-\beta \Delta A_{bound}}}{e^{-\beta \Delta A_{unbound}}} = e^{-\beta \Delta \Delta A} \qquad (4)$$

For convenience, we will discuss the allosteric efficacy in terms of the quantity $\Delta\Delta A$, which we will call the *thermodynamic coupling*.

$\Delta\Delta A$ is symmetric at equilibrium, e.g. $\Delta\Delta A$ for receptor activation conditional on ligand binding is equivalent to the $\Delta\Delta A$ for ligand binding conditional on activation. Thus, the following two definitions of $\Delta\Delta A$ are equivalent:

$$\Delta\Delta A = \Delta A\left(R_{on} + L \rightarrow R_{on}L\right) - \Delta A\left(R_{off} + L \rightarrow R_{off}L\right)$$
$$\Delta\Delta A = \Delta A\left(R_{off}L \rightarrow R_{on}L\right) - \Delta A\left(R_{off} + L \rightarrow R_{on} + L\right)$$

$$(5)$$

However, there is no reason to assume that the receptor activation is a two-state process. In fact, NMR experiments have revealed a multi-modal activation process in the $\beta_2$-adrenergic receptor $(\beta_2 AR)$[4], and quantitative mass spectroscopy experiments have revealed ligand-specific states in the same system[5]. These results along with other evidence for additional states in $\beta_2 AR$ and other receptors[6–11] indicate that activation must be treated as either involving more than two discrete states, or even as involving a continuous conformational space.

Receptor activation involves not only multiple states, but also multiple dimensions. The complex behavior of an allosteric receptor is thus unlikely to be well described by a single reaction coordinate; instead, the large number of potential conformational states may be best described by multiple collective variables (CVs; variables that are functions of the atomic coordinates) that are thermodynamically coupled in non-trivial ways. At the very minimum, it is impossible to understand the molecular mechanism of ligand-induced receptor activation without explicitly considering the thermodynamic coupling between the ligand binding site and active site, so that a minimal set of CVs should at least include one CV for each one of these sites. These CVs are considered intrinsic to the specific receptor, and their thermodynamic coupling arises from the complex molecular interaction network that separates them spatially in the receptor's structure. This intrinsic thermodynamic coupling is of great interest, as an understanding of the nature of this coupling can be used to inform both the design of ligands that modulate function in a highly specific manner, and the design of receptors with modified allosteric properties. We have previously represented this intrinsic thermodynamic coupling using the recently developed Allosteric Ising Model (AIM)[12], a two-state model of allostery that implicitly includes the potential energy of interaction between structural components. While the AIM and other statistical mechanical models of allostery, such as the ensemble allosteric model[13], have allowed us to derive some analytical features of simple allosteric systems, a general method that does not rely on the two-state assumption is still needed to study the intrinsic thermodynamic coupling between structural components in real systems. Here we describe such a method and illustrate its capabilities by showing how allosteric coupling in the alanine dipeptide molecule contributes to the overall shape of the $\Phi/\Psi$ free energy surface, and by identifying the interactions that are necessary for this coupling and their contributions to the energetics.

## Theoretical Developments

### Derivation

To quantify the intrinsic thermodynamic coupling between CVs, we will derive expressions analogous to the allosteric efficacy for the coupled perturbation of discrete or continuous CVs. away from their equilibrium distributions. Let $\vec{r} \in \mathbb{R}^N$ represent the coordinates of the allosteric protein and its environment that define our system, which does not include any ligand that we consider here as external perturbations. The probability density of each microstate $\vec{r}$ is given by the Boltzmann distribution,

$$f(\vec{r}) = \frac{e^{-\beta U(\vec{r})}}{\int e^{-\beta U(\vec{r})}\, d\vec{r}} \ , \tag{6}$$

where $U(\vec{r})$ is the potential energy function. The numerator is the Boltzmann factor denoted as

$$B(\vec{r}) = e^{-\beta U(\vec{r})} \ . \tag{7}$$

The free energy can be written as a functional of the Boltzmann factor function,

$$A\left[B(\vec{r})\right] = -\frac{1}{\beta} \log\left(\int e^{-\beta U(\vec{r})}\, d\vec{r}\right) \ . \tag{8}$$

We define a CV as a function $X(\vec{r})$ of the system's coordinates that can be either continuous or discrete. For a continuous CV, the probability density function is

$$f(x) = \frac{\int \delta\left(X(\vec{r}) - x\right) e^{-\beta U(\vec{r})}\, d\vec{r}}{\int e^{-\beta U(\vec{r})}\, d\vec{r}} \ . \tag{9}$$

For a discrete CV, the probability mass function, $p(x)$, is defined by an identical expression, but is bounded to be $< 1$ everywhere. We can calculate the free energy conditional on a value of the CV as

$$A\left[B(\vec{r})|X(\vec{r}) = x\right] = -\frac{1}{\beta} \log\left(\int \delta\left(X(\vec{r}) - x\right) e^{-\beta U(\vec{r})}\, d\vec{r}\right) \ . \tag{10}$$

Equation (10) can be rewritten in terms of either $f(x)$ or $p(x)$. Because we use the histogram method to estimate the probability mass function of the CVs in the application following this derivation, we will assume discrete CVs described using $p(x)$ without loss of generality. The free energy becomes

$$A\left[B(\vec{r})|X(\vec{r}) = x\right] = -\frac{1}{\beta} \log\left(p(x)\right) + A\left[B(\vec{r})\right] \ . \tag{11}$$

Consider a second CV $Y(\vec{r})$, with analogous probability function and free energy definitions. A joint probability mass function for the two CVs can be written as

$$p(x,y) = \frac{\int \delta\big(X(\vec{r})-x\big)\delta\big(Y(\vec{r})-y\big)e^{-\beta U(\vec{r})}\,d\vec{r}}{\int e^{-\beta U(\vec{r})}\,d\vec{r}} \ , \tag{12}$$

so that the analogous free energy conditional on values of both CVs is

$$A\Big[B\big(\vec{r}\,|\,X(\vec{r})=x,Y(\vec{r})=y\big)\Big] = -\frac{1}{\beta}\log\big(p(x,y)\big) + A\big[B(\vec{r})\big] \ . \tag{13}$$

One can imagine $X(\vec{r})$ to describe the ligand binding site and $Y(\vec{r})$ to describe the active site of the protein; the binding of a ligand to the system then acts as an external perturbation to the distributions of these CVs. To quantify the intrinsic coupling between these CVs, we apply artificial perturbations to the equilibrium CV distributions such that one or both CVs become constrained to a given value. From the equilibrium state and these artificially perturbed states, we calculate the allosteric efficacy of the following thermodynamic cycle:

$$
\begin{array}{ccc}
f(\vec{r}) & \xleftarrow{\ \Delta A_1\ } & f\big(\vec{r}\,|\,X(\vec{r})=x\big) \\[4pt]
\Big\uparrow{\scriptstyle \Delta A_4} & \begin{array}{c}\Delta\Delta A = \Delta A_2 - \Delta A_1 \\ \Delta\Delta A = \Delta A_3 - \Delta A_4\end{array} & \Big\uparrow{\scriptstyle \Delta A_3} \\[4pt]
f\big(\vec{r}\,|\,Y(\vec{r})=y\big) & \xleftarrow{\ \Delta A_2\ } & f\big(\vec{r}\,|\,X(\vec{r})=x,Y(\vec{r})=y\big)
\end{array}
\tag{14}
$$

We will refer to this class of thermodynamic cycles as "thermodynamic perturbation cycles". The thermodynamic coupling of the perturbations at position $(x,y)$ in the CV space, $\Delta\Delta A(x,y)$, can be calculated as

$$\Delta\Delta A(x,y) = A\Big[B\big(\vec{r}\,|\,X(\vec{r})=x,Y(\vec{r})=y\big)\Big] - A\Big[B\big(\vec{r}\,|\,X(\vec{r})=x\big)\Big] - A\Big[B\big(\vec{r}\,|\,Y(\vec{r})=y\big)\Big] + A\big[B(\vec{r})\big] \ . \tag{15}$$

Equation (15) simplifies to

$$\Delta\Delta A(x,y) = -\frac{1}{\beta}\log\left(\frac{p(x,y)}{p(x)p(y)}\right) \ . \tag{16}$$

This is the mathematical definition we propose for the central quantity $\Delta\Delta A(x,y)$ that we call the *thermodynamic coupling function* for the CVs $X(\vec{r})$ and $Y(\vec{r})$. In two dimensions, Eq. (16) defines what we call the *allostery landscape* (see Fig. 1 for an example).

It should be noted that the thermodynamic coupling function has a natural normalization when the CVs are discrete. If the two CVs are maximally coupled, constraining one CV will fully constrain the other. Thus, at maximum coupling,

$$A\Big[B\big(\vec{r}\,|\,X(\vec{r})=x\big)\Big]=A\Big[B\big(\vec{r}\,|\,Y(\vec{r})=y\big)\Big]=A\Big[B\big(\vec{r}\,|\,X(\vec{r})=x,Y(\vec{r})=y\big)\Big]\,, \qquad \textbf{(17)}$$

and thus

$$\Delta\Delta A_{max}\big(x,y\big)=A\Big[B(\vec{r})\Big]-A\Big[B\big(\vec{r}\,|\,X(\vec{r})=x,Y(\vec{r})=y\big)\Big]\,. \qquad \textbf{(18)}$$

We can then normalize (16) to this upper bound to define the normalized allosteric coupling (AC),

$$AC(x,y)=-\frac{\Delta\Delta A(x,y)}{\Delta\Delta A_{max}(x,y)}=\frac{\log\big(p(x)p(y)\big)}{\log\big(p(x,y)\big)}-1\,. \qquad \textbf{(19)}$$

The AC ranges from 1 to -1 and matches the convention commonly used for positive and negative allostery; positive values indicate that constraining one CV reduces the free energy required to constrain the other, whereas negative values indicate that constraining one CV increases the free energy required to constrain the other. In essence, *the magnitude of the AC describes what fraction of the maximal allostery is contributing to the free energy of the joint state, whereas the sign of the AC describes whether that allostery is positive or negative*.

When applied to the biophysical information transmission process occurring in a receptor (i.e. the thermodynamic coupling between the ligand binding site and the active site), the definitions above indicate that the thermodynamic coupling function is negative if measuring the active site to be in the "active" state reduces the uncertainty associated with whether or not the ligand binding site is in the "bound" state. The two-state ligand-induced receptor activation model defined in Eq. (1) can be just as easily described using the thermodynamic coupling function if the collective variables $X(\vec{r})$ and $Y(\vec{r})$ are defined to take only two discrete values (bound/unbound and on/off, respectively). In this context, the two-state allosteric efficacy in Eq. (4) can be calculated from an allosteric cycle composed of four allosteric perturbation cycles:

$$
\begin{array}{ccccc}
f\left(\vec{r}\,|L_{unbound},R_{off}\right) & \rightleftharpoons & f\left(\vec{r}\,|L_{unbound}\right) & \rightleftharpoons & f\left(\vec{r}\,|L_{unbound},R_{on}\right) \\
\rightleftharpoons & & \rightleftharpoons & & \rightleftharpoons \\
f\left(\vec{r}\,|R_{off}\right) & \rightleftharpoons & f\left(\vec{r}\right) & \rightleftharpoons & f\left(\vec{r}\,|R_{on}\right) \\
\rightleftharpoons & & \rightleftharpoons & & \rightleftharpoons \\
f\left(\vec{r}\,|L_{bound},R_{off}\right) & \rightleftharpoons & f\left(\vec{r}\,|L_{bound}\right) & \rightleftharpoons & f\left(\vec{r}\,|L_{bound},R_{on}\right)
\end{array}
\qquad \textbf{(20)}
$$

so that

$$
\Delta\Delta A_{cycle} = \Delta\Delta A\left(L_{bound},R_{on}\right) + \Delta\Delta A\left(L_{unbound},R_{off}\right) - \Delta\Delta A\left(L_{unbound},R_{on}\right) - \Delta\Delta A\left(L_{bound},R_{off}\right) \;. \qquad \textbf{(21)}
$$

Thus, a large negative $\Delta\Delta A_{cycle}$ for the allosteric coupling between activation and ligand binding indicates that when the receptor is in the active state, the uncertainty that a ligand is bound is greatly reduced, whereas when the receptor is in the inactive state, the uncertainty that a ligand is not bound is greatly reduced.

**Relationships to the mutual information and the copula**

The mutual information, which is often used to quantify allostery[14–17], is defined as

$$
I_2\left[p\left(x,y\right)\right] = \sum_{x \in X}\sum_{y \in Y} p\left(x,y\right)\log\left(\frac{p\left(x,y\right)}{p\left(x\right)p\left(y\right)}\right) \;. \qquad \textbf{(22)}
$$

Interestingly, $\Delta\Delta A(x,y)$ in (16) is proportional to the argument of the logarithm in (22), which is known as the pointwise mutual information (PMI)[18] [REF],

$$PMI(x,y) = \log\left(\frac{p(x,y)}{p(x)p(y)}\right).$$ **(23)**

Like the allosteric efficacy, the PMI is symmetric (i.e. the order of variables x and y does not matter). To understand the PMI from the perspective of information theory, one can consider the information gained due to the reduction in uncertainty associated with measuring a variable. This information gain by measuring $X(\vec{r})$ to be equal to x is

$$I(x) = -\log\left(p(x)\right).$$ **(24)**

However, if two variables are measured, and those variables are dependent on each other, the amount of information gained by measuring the second variable will be different than if it was measured alone. For example, if $Y(\vec{r})$ was measured to be y, the probability distribution of $X(\vec{r})$ changes, and thus the information gained by measuring $X(\vec{r})$ is now

$$I(x|y) = -\log\left(p(x|y)\right).$$ **(25)**

The PMI is the difference in the information gain,

$$PMI(x,y) = I(x) - I(x|y).$$ **(26)**

The mutual information is the PMI weighted by the joint probability density function. Consequently, the mutual information gives a high weight to the thermodynamic coupling of perturbations of high equilibrium probability states and low weight to those of low equilibrium probability. This is important for the mechanistic interpretation of allosteric couplings that are quantified only by their mutual information, as functionally significant perturbations do not necessarily drive the protein towards a region of its intrinsic CV space that is already high probability prior to perturbation. In fact, perturbations such as ligands generally drive the system away from the unbound equilibrium (e.g. where the inactive state is preferred to the active state), so the mutual information would end up giving larger weight to less functionally relevant states.

In such cases, when considering only the protein's degrees of freedom, the mutual information is not a good quantification of the intrinsic thermodynamic couplings that mediate the system's response to ligand binding. Even in the simplest case of the allosteric coupling between ligand binding and activation as described in the ATSM, the mutual information between ligand binding and activation will depend on the affinity of the ligand, and will go to 0 as the affinity goes to either 0 or $\infty$, independent of the allosteric efficacy of the ligand. Therefore, we argue that it becomes preferable instead to analyze the entire 2-dimensional thermodynamic coupling surface, $\Delta\Delta A(x,y)$, which we call the "allostery landscape", as it contains information regarding the allosteric efficacy for all possible perturbations to the distribution of those CVs.

The thermodynamic coupling function is also related to the copula density function from probability theory[19]. The copula density function of a bivariate probability distribution is

$$c(x,y) = \frac{p(x,y)}{p(x)p(y)} \ . \tag{27}$$

Any multivariate distribution can be expressed as a set of marginal probability distributions and a copula that defines the dependency between them[19], and the entropy of the copula distribution is equivalent to the mutual information[20]. Thus, a multivariate thermodynamic coupling function behaves like a copula, defining the information transmission properties of the allosteric system. The relationship between the thermodynamic coupling function and fundamental concepts in information theory and probability theory suggest that past work in these fields may be able to be adapted for biophysical applications and provide new insights into allostery.

**Contribution of specific energy terms**

Having introduced a quantification of the allosteric coupling between two CVs with the allostery landscape, a major mechanistic question still remains. What features of the structure and energetics of a given system define the thermodynamic coupling function? To answer this question, we derive the change in thermodynamic coupling function when a generic biasing potential energy term $U_{bias}(\vec{r})$ is added to the system's total potential energy function, $U_{bias}(\vec{r})$. The change in thermodynamic coupling (16) at any point in the CV space can be estimated using a free energy perturbation approach, which we will refer to here as "biasing" to avoid confusion with "perturbing" that refers to constraining the system at (x,y) in the CV space in Eq. (14). The change in free energy of the system when a biasing potential is added is

$$\Delta A\Big[f(\vec{r}) \rightarrow f_{bias}(\vec{r})\Big] = -\frac{1}{\beta}\log\Big(\int e^{-\beta U_{bias}(\vec{r})}f(\vec{r})d\vec{r}\Big) \ . \tag{28}$$

The change in free energy of the perturbed states can be similarly written as

$$\Delta A\Big[f(\vec{r}|x,y) \rightarrow f_{bias}(\vec{r}|x,y)\Big] = -\frac{1}{\beta}\log\Big(\int e^{-\beta U_{bias}(\vec{r})}f(\vec{r}|x,y)d\vec{r}\Big) \ . \tag{29}$$

Thus, the biased thermodynamic coupling function is

$$\Delta\Delta A_{bias}(x,y) = \Delta\Delta A(x,y) - \frac{1}{\beta}\log\left(\frac{\int e^{-\beta U_{bias}(\vec{r})}f(\vec{r}|x,y)d\vec{r}\int e^{-\beta U_{bias}(\vec{r})}f(\vec{r})d\vec{r}}{\int e^{-\beta U_{bias}(\vec{r})}f(\vec{r}|x)d\vec{r}\int e^{-\beta U_{bias}(\vec{r})}f(\vec{r}|y)d\vec{r}}\right) \ . \tag{30}$$

We wish to understand the contribution to the thermodynamic coupling function of a structural feature of interest, or of a specific interaction between structural elements of the system. Assuming this feature of interest is described by a specific energy term $U_{int}(\vec{r})$ of the total potential energy function $U(\vec{r})$, we can use Eq. (30) with a biasing potential that is equal and opposite to that energy term, $U_{bias}(\vec{r}) = -U_{int}(\vec{r})$. In the next section, we use this approach to quantify the contribution of specific interactions in the alanine dipeptide system by mapping the corresponding change in thermodynamic coupling, $\Delta\Delta\Delta A(x,y) = \Delta\Delta A(x,y) - \Delta\Delta A_{bias}(x,y)$.

Consider the special case of a particular energy term of interest that is a function of a collective variable $Z(\vec{r})$. This corresponds to a biasing potential of $U_{bias}(\vec{r}) = -U_{int}(Z(\vec{r}))$. We have found (see Appendix) that this result in the following biased thermodynamic coupling functions:

1. If $Z(\vec{r})$ is independent of either $X(\vec{r})$ or $Y(\vec{r})$,

$$\Delta\Delta A_{bias}(x,y) = \Delta\Delta A(x,y) .$$

2. If $Z(\vec{r})$ is conditionally independent of $Y(\vec{r})$ given $X(\vec{r})$, or if $Z(\vec{r}) = X(\vec{r})$,

$$\Delta\Delta A_{bias}(x,y) = \Delta\Delta A(x,y) - \frac{1}{\beta}\log\left(\frac{\int e^{\beta U(Z(\vec{r}))}f(\vec{r})d\vec{r}}{\int e^{\beta U(Z(\vec{r}))}f(\vec{r}|y)d\vec{r}}\right) . \qquad (31)$$

The second result is interesting because in this case $\Delta\Delta\Delta A(x,y)$ becomes a function of y only. A corresponding result is found if $X(\vec{r})$ and $Y(\vec{r})$ are permuted.

Importantly, these findings indicate that the influence of any specific energy term on the thermodynamic coupling between two CVs can be clearly defined from the difference between the unbiased and biased thermodynamic coupling functions. The two conditions described above indicate that unbiasing by a potential energy term that mediates an allosteric coupling will have a two-dimensional effect on the thermodynamic coupling function (i.e. the second term in (30) is dependent on both $X(\vec{r})$ and $Y(\vec{r})$). We know that, due to non-additive effects, the free energy contributions of such coupled energy terms cannot be rigorously deconvoluted[21]. Therefore, the contributions obtained by biasing the allosteric coupling function (shown in Fig. 2), cannot be taken as a *sensu stricto* decomposition of $\Delta\Delta A(x,y)$. This method nonetheless allows for the detailed analysis of the mechanism of allosteric coupling and can applied generally across any system whose conformational ensemble can be sampled using methods such as molecular dynamics (MD).

# Application: The Alanine Dipeptide

### Estimating the thermodynamic coupling function

To illustrate the use and utility of the thermodynamic coupling function, we analyzed the allostery landscape of the alanine dipeptide in solution. The alanine dipeptide free energy landscape is a popular model system for testing enhanced sampling and free energy methods as the entire system can be described well by only two CVs, the $\Phi$ and $\Psi$ torsion angles along the bonds connecting the alanine $C_\alpha$ atom to the capped N- and C-terminus, respectively (see Fig. 1a). Here, in analogy to larger allosteric systems, we consider that $\Phi$ captures the state of the N-terminal domain and $\Psi$ the state of the C-terminal domain, and we ask the question of how the N-terminal and C-terminal domains of the protein are allosterically coupled. Despite the small size of the system, the irregular features of the $\Phi/\Psi$ free energy surface indicates that these CVs are thermodynamically coupled in a non-trivial way, and thus the alanine dipeptide is an ideal model system for illustrating the power of the thermodynamic coupling function.

We constructed the 2-dimensional $\Phi/\Psi$ probability density function of alanine dipeptide in water from five independent 50 ns trajectories produced with driven adiabatic free energy dynamics[22,23] (see Methods). Following a protocol that we previously demonstrated to yield well-converged free energy surfaces up to 40 kJ/mol above the global minimum[24,25], we reconstructed the free energy landscape shown in Fig. 1b using the reweighted histogram estimator. In order to investigate which features of the alanine dipeptide free energy landscape are due to thermodynamic coupling between the two angles, we used Eq. (16) to calculate the thermodynamic coupling function shown in Fig. 1c. Significant allosteric couplings are evident in the regions of the left-handed α-helix (known as $\alpha_L$, which should not be confused with the symbol for allosteric efficacy) and the $C_{7ax}$, indicating that if Φ is driven to the (0 to 2 rad / 0° to 120°) region, the transition of Ψ to the (0 to 2 rad / 0° to 120°) and (-1 to -2 rad / -60° to -180°) regions becomes more favorable.

In the normalized AC landscape of the alanine dipeptide, calculated according to Eq, (19) and shown in Fig. 1d, the $\alpha_L$ and $C_{7ax}$, regions have couplings of around 0.4, indicating that a substantial amount of the maximal theoretically possible Φ/Ψ allostery contributes to these state's stabilities. Thus, while these regions have a relatively low probability, our analysis suggests that the allosteric coupling accounts for the small but significant populations of $\alpha_L$ and $C_{7ax}$ conformations that appear at equilibrium. In addition, there appears to be significant coupling present at the transition region between $\alpha_L$ and $C_{7eq}$, and to a lesser extent between $\alpha_R$ and $C_{7ax}$, indicating that these transitions may also be facilitated by allostery. We also see significant unfavorable allosteric coupling in the high free energy regions, which indicates that a thermodynamic coupling between the CVs contributes to the high free energy of these regions.

The mutual information, Eq. (22), between Φ and Ψ is 1.11±0.01 nats (95% confidence interval from bootstrapping), or for better comparison to the thermodynamic coupling, 0.29±0.03 kJ/mol at 300 K. It should be noted that these values are quite low compared to the numerous regions of high thermodynamic coupling and normalized AC ($ abs\left(\Delta\Delta A\left(x,y\right)\right) >$ 6 kJ/mol, abs(AC(x,y)) > 0.3, see Fig. 1), and thus utilizing the mutual information alone understates the thermodynamic coupling between Φ and Ψ. Mapping the quantity summed over in Eq. (22) (see

Fig. S1) shows that the major contributions to the mutual information come from very localized regions of the CV space. Thus, using a single number to quantify the coupling between $\Phi$ and $\Psi$ misses the fact that the allostery landscape has significant regions of both negative and positive coupling. This can be important if, for example, one seeks to design a ligand that allosterically stabilizes a lower probability state.

**The influence of specific interactions on the allosteric coupling between termini**

In order to understand which structural features contribute to the thermodynamic coupling of $\Phi$ and $\Psi$, we decomposed the alanine dipeptide into three structural components: i) the N-terminus, which includes all atoms on the N-terminal side of the $C_\alpha$ carbon, ii) the C-terminus, which includes all atoms on the C-terminal side of $C_\alpha$, and iii) the "channel", which includes $C_\alpha$ as well as the hydrogen and methyl side chain bound to it. These three structural components can mediate the $\Phi/\Psi$ thermodynamic coupling through three different mechanisms: i) direct non-bonded interaction of the termini (estimated with a dielectric constant $\varepsilon = 60$), ii) indirect interaction of the termini through non-bonded interactions with the channel, and iii) indirect interaction of the termini through bonded interaction with the channel. Thus, we estimated the potential energy $U_{int}(\bar{r})$ contributed by each of these groups of energy terms for each frame along the trajectories and reweighted the free energy landscape with an equal and opposite biasing potential $U_{bias}(\bar{r})$ according to Eq. (28), see Figs. S2-S4. We then calculated the contribution of $U_{int}(\bar{r})$ to the thermodynamic coupling landscape $\Delta\Delta A(x,y)$ using Eqs. (29) and (30), as represented on Fig. 2 in the form of $\Delta\Delta\Delta A(x,y) = \Delta\Delta A(x,y) - \Delta\Delta A_{bias}(x,y)$. As a control, we also unbiased using the potential energy term corresponding exactly to the definition of the $\Phi(\bar{r})$ CV, i.e. the C-N-$C_\alpha$-C dihedral angle. This reweighting resulted in one-dimensional variations of the thermodynamic coupling function along the $\Psi$ axis, as expected from Eq. (31) (see Fig 2a).

Interestingly, while chemical intuition may suggest that the direct interaction of the termini is the major mediator of the thermodynamic coupling, we find that the direct non-bonded interaction only contributes to the negative thermodynamic coupling surrounding the central forbidden region, as show in Fig. 2b. The only other significant change to the thermodynamic coupling function are one-dimensional bands at $\Psi \sim 1$ and $\Psi \sim -1$ rad, which indicates that the interaction itself my to some extent be indirectly coupled to $\Phi$ through its direct dependency on $\Psi$.

Fig. 2d shows that the bonded interactions between the termini and the channel are the most significant contributors to both the positive and negative thermodynamic coupling between the termini, while the non-bonded interactions between the termini and the channel (Fig. 2c) do not significantly contribute to the thermodynamic coupling. These results suggest that $\Phi$ and $\Psi$ become thermodynamically coupled due to the energetics of the bonds, angles, and dihedrals composed of atoms shared between each terminus and the channel. For example, the $\Phi$ and $\Psi$ dihedrals each share the angle formed by three central atoms. Different combinations of $\Phi$ and $\Psi$ frustrate this central angle to different extents, leading to a thermodynamic coupling between the two. We however note that in the alanine dipeptide system, the energy terms described above are tightly coupled with each other, as well as will other energy terms (such as the internal bonded energy of the channel). Therefore, the contributions represented in Fig. 2 do not represent an exact decomposition of $\Delta\Delta A(x,y)$ and must be regarded as useful cues for the qualitative understanding of how allosteric coupling can be established between two domains of a molecular system.

## Conclusions

We have derived a thermodynamic coupling function based on the allosteric efficacy that quantifies the allosteric coupling between two continuous or discrete CVs. We find that the thermodynamic coupling function is related to the pointwise mutual information and the copula, and is best represented in the form of an allostery landscape, in units of free energy. Such a representation reveals the allosteric response to all possible perturbations of the CVs. We showed that the allostery landscape of the $\Phi$ and $\Psi$ dihedral angles of the alanine dipeptide's contains positive allosteric couplings that appear to stabilize the $\alpha_L$ and $C_{7ax}$ conformations, and negative allosteric couplings that coincide with the high free energy regions of the $\Phi/\Psi$ space. Based on the formalism we developed, we were able to attribute features of this thermodynamic coupling landscape to specific interaction energy terms, thus allowing interpretation of the allosteric landscape. It is important to note that the criterion introduced here for determining whether a specific interaction mediates an allosteric coupling is more rigorous than our previous n-body information-based criterion[14]. While the 3-body information between three CVs is in fact a function of unbiased and biased thermodynamic coupling functions (see Appendix, Eq. (46)), if $Z(r)$ is conditionally independent of $X(r)$ or $Y(r)$ given the other CV, the 3-body information will be maximal. Consequently, the 3-body information does not permit to determine definitively whether $Z(r)$ mediates a thermodynamic coupling between $X(r)$ and $Y(r)$, or if one of the CVs mediates a thermodynamic coupling between $Z(r)$ and the other. Specifically, the 3-body information criterion will include some number of false positives (as we have previously described[14]), whereas all structural features that correspond to a potential energy term and have a two-dimensional influence on the thermodynamic coupling function can be considered to be effective mediators of the thermodynamic coupling.

The concepts developed here are very general and are applicable to larger molecular systems, provided enough sampling is available. This method can become a powerful tool in understanding the molecular mechanisms of the many proteins in which allostery is essential to biological function, as it has the potential to identify novel allosteric sites that modulate functionally important reaction coordinates. Such capabilities should be very useful in the design of novel therapeutics that allosteric modulate their specific targets in new ways, as well as for the detailed analysis of allosteric mechanisms than can guide the design of the synthetic allosteric proteins.

## Methods

The alanine dipeptide (N-Acetyl-Alanine-N'-Methyl amide) was modeled with the all-atom charmm22* force field[26] and solvated in explicit TIP3P water molecules[27]. charmm22* was chosen as it is able to reproduce an accurate alanine dipeptide free energy landscape without utilizing the CMAP[28] term used by other force fields. We chose to avoid force fields using the CMAP term as it induces a trivial thermodynamic coupling through a direct interaction between $\Phi$ and $\Psi$, rather than allowing it to emerge from separate terms of the traditional potential energy function. Molecular dynamics simulation were performed using the Charmm port[29] in the Gromacs 4.5 program[30] with particle-mesh Ewald[31] treatment of electrostatics and Lennard-Jones interactions switched off between 10Å and 12Å.

The systems were maintained at temperature T=300K with Nosé-Hoover chain thermostats[32]. Similarly to our previous study on dipeptides[25], enhanced sampling was achieved with driven adiabatic free energy dynamics[22,23] (dAFED), also known as temperature accelerated molecular dynamics[33] (TAMD), implemented in the PLUMED plugin[34]. Two collective variables (CVs), defined as the backbone dihedral angles $\Phi$ and $\Psi$, were coupled (harmonic constant 1000 kJ/mol/rad$^2$) to heavy fictitious particles (pseudo-mass 50 amu•nm$^2$/rad$^2$) held at temperature $T_s = 600$ K by generalized Gaussian Moment thermostats (order 2)[35]. Simulations were conducted in five independent replicates of 50 ns each after a standard equilibration phase starting with independent initial velocities. Free energy surfaces (FESs) in the ($\Phi,\Psi$) plane were reconstructed[24] using the reweighted histogram smoothed with multivariate Gaussian kernel regression in Matlab (release 2015b, The MathWorks, Inc., Natick, Massachusetts, United States). A cutoff of 40 kJ/mol was used for the FESs, above which sampling was too poor for reliable surface estimation.

In principle, estimating an observable from a dAFED/TAMD simulation requires binning the observable values in the CV space, and reweighting each bin by a function of the FES at this point[36]. However, $\Delta\Delta A(x,y)$ in Eq. (16) depends only on the probability mass at 300K in the CV space, $p(\Phi,\Psi)$. This can be derived directly from the density obtained from the dAFED/TAMD simulation, $p_{adb}(\Phi,\Psi)$, by rescaling and re-normalizing,

$$p(\Phi,\Psi) \propto p_{adb}(\Phi,\Psi)^{\frac{T_s}{T}} \qquad (32)$$

Due to the surface smoothing steps, propagation of uncertainties is not practical for estimating confidence intervals on the allostery landscape. Instead, we use the bootstrapping approach[37]. Specifically, because observations from MD time series are notoriously not independent, we use block bootstrapping[38], i.e. we generate artificial samples by drawing at random (with replacement) segments of trajectory of 1 ns in length. Then, for each bin in the $(\Phi, \Psi)$ plane, we estimate a 95% confidence interval for the allosteric coupling function and for the AC based on the standard deviation among the bootstrapped samples. If in a given bin this confidence interval includes the value zero, the existence of an allosteric effect cannot be assessed with certainty in this bin and we represent it in a greyed-out color in panels (c) and (d) of Fig. 2 and in Fig. 3.

## Appendix

**Biasing the thermodynamic coupling function with potentials that are functions of a CV**

If $X(\bar{r})$ and $Z(\bar{r})$ are independent,

$$p(z|x) = p(z) \ .$$
(33)

Thus, we can rewrite the integrals in the biased thermodynamic coupling (30) as:

$$\int e^{-\beta U_{bias}(\bar{r})} p(\bar{r}|x,y) d\bar{r} = \int e^{\beta U_{im}(z)} p(z|x,y) dz = \int e^{\beta U_{im}(z)} p(z|y) dz$$

$$\int e^{-\beta U_{bias}(\bar{r})} p(\bar{r}|y) d\bar{r} = \int e^{\beta U_{im}(z)} p(z|y) dz$$

$$\int e^{-\beta U_{bias}(\bar{r})} p(\bar{r}|x) d\bar{r} = \int e^{\beta U_{im}(z)} p(z|x) dz = \int e^{\beta U_{im}(z)} p(z) dz$$

$$\int e^{-\beta U_{bias}(\bar{r})} p(\bar{r}) d\bar{r} = \int e^{\beta U_{im}(z)} p(z) dz \ .$$
(34)

and we find that

$$\Delta\Delta A_{bias}(x,y) = \Delta\Delta A(x,y)$$
(35)

The equivalent is true if $Y(\bar{r})$ and $Z(\bar{r})$ are independent.

If $X(\bar{r})$ and $Z(\bar{r})$ are independent given $Y(\bar{r})$,

$$p(z|x,y) = p(z|y) \ ,$$
(36)

we can rewrite the integrals in the biased thermodynamic coupling (30) as:

$$\int e^{-\beta U_{bias}(\vec{r})}p(\vec{r}|x,y)d\vec{r} = \int e^{\beta U_{im}(z)}p(z|x,y)dz = \int e^{\beta U_{im}(z)}p(z|y)dz$$

$$\int e^{-\beta U_{bias}(\vec{r})}p(\vec{r}|y)d\vec{r} = \int e^{\beta U_{im}(z)}p(z|y)dz$$

$$\int e^{-\beta U_{bias}(\vec{r})}p(\vec{r}|x)d\vec{r} = \int e^{\beta U_{im}(z)}p(z|x)dz \qquad \textbf{(37)}$$

$$\int e^{-\beta U_{bias}(\vec{r})}p(\vec{r})d\vec{r} = \int e^{\beta U_{im}(z)}p(z)dz \ .$$

Thus,

$$\Delta\Delta A_{bias}(x,y) = \Delta\Delta A(x,y) - \frac{1}{\beta}\log\left(\frac{\int e^{\beta U_{im}(z)}p(z)dz}{\int e^{\beta U_{im}(z)}p(z|x)z}\right) = \Delta\Delta A(x,y) - \frac{1}{\beta}\log\left(\frac{\int e^{\beta U_{im}(Z(\vec{r}))}p(\vec{r})d\vec{r}}{\int e^{\beta U_{im}(Z(\vec{r}))}p(\vec{r}|x)d\vec{r}}\right) \ . \qquad \textbf{(38)}$$

If $Y(\vec{r})$ and $Z(\vec{r})$ are independent given $X(\vec{r})$, we can use a similar simplification to find:

$$\Delta\Delta A_{bias}(x,y) = \Delta\Delta A(x,y) - \frac{1}{\beta}\log\left(\frac{\int e^{\beta U(Z(\vec{r}))}p(\vec{r})d\vec{r}}{\int e^{\beta U(Z(\vec{r}))}p(\vec{r}|y)d\vec{r}}\right) \ . \qquad \textbf{(39)}$$

**The three-body information**

The three-body information shared by three CVs $X(\vec{r})$, $Y(\vec{r})$, and $Z(\vec{r})$ is defined as

$$I_3\big[p(x,y,z)\big] = I_2\big[p(x,y)\big] - I_2\big[p(x,y|z)\big] \ . \qquad \textbf{(40)}$$

This can be expanded to

$$I_3\big[p(x,y,z)\big] = \sum_{x\in X}\sum_{y\in Y}p(x,y)\Delta\Delta A(x,y) - \sum_{z\in Z}\sum_{x\in X}\sum_{y\in Y}p(x,y,z)\log\left(\frac{p(x,y|z)}{p(x|z)p(y|z)}\right) \ . \qquad \textbf{(41)}$$

The second term can rewritten as

$$I_2\big[p(x,y|z)\big] = \sum_{z\in Z}\sum_{x\in X}\sum_{y\in Y}p(x,y,z)\log\left(\frac{p(x,y,z)p(z)}{p(x,z)p(y,z)}\right) \ , \qquad \textbf{(42)}$$

and then expanded to

$$I_2\big[p(x,y|z)\big] = \sum_{z\in Z}\sum_{x\in X}\sum_{y\in Y}p(x,y,z)\log\left(\frac{\int\delta(Z(\vec{r})-z)p(\vec{r}|x,y)d\vec{r}\int\delta(Z(\vec{r})-z)p(\vec{r})d\vec{r}}{\int\delta(Z(\vec{r})-z)p(\vec{r}|x)d\vec{r}\int\delta(Z(\vec{r})-z)p(\vec{r}|y)d\vec{r}}\right) \ . \qquad \textbf{(43)}$$

The Dirac delta functions can be rewritten as harmonic biasing potential terms with strictly positive force constants, $k$

$$\delta\left(Z(\vec{r})-z\right) = e^{-U_{bias}(Z(\vec{r})-z)} = \lim_{k\to\infty}\left[e^{-\frac{\beta k}{2}(Z(\vec{r})-z)^2}\right],\qquad \textbf{(44)}$$

such that Eq. (43) becomes

$$I_2\left[p(x,y|z)\right] = \sum_{z\in Z}\sum_{x\in X}\sum_{y\in Y}p(x,y,z)\log\left(\frac{\int e^{U_{bias}(Z(\vec{r})-z)}p(\vec{r}|x,y)d\vec{r}\int e^{U_{bias}(Z(\vec{r})-z)}p(\vec{r})d\vec{r}}{\int e^{U_{bias}(Z(\vec{r})-z)}p(\vec{r}|x)d\vec{r}\int e^{U_{bias}(Z(\vec{r})-z)}p(\vec{r}|y)d\vec{r}}\right) = \sum_{z\in Z}\sum_{x\in X}\sum_{y\in Y}p(x,y,z)\Delta\Delta A_{bias}(x,y)\ . \qquad \textbf{(45)}$$

Thus,

$$I_3\left[p(x,y,z)\right] = \sum_{x\in X}\sum_{y\in Y}p(x,y)\Delta\Delta A(x,y) - \sum_{z\in Z}\sum_{x\in X}\sum_{y\in Y}p(x,y,z)\Delta\Delta A_{bias(z)}(x,y)\ . \qquad \textbf{(46)}$$

This shows that the 3-body information is a function of both the original thermodynamic coupling function and the biased thermodynamic coupling function for a fixed value of $Z(\vec{r})$.

# Acknowledgements

The authors gratefully acknowledge support from the National Institute of Health grants P01 DA012408 and U54 GM087519. Simulations were performed at the Vital-IT High Performance Computing Center of the Swiss Institute of Bioinformatics, and free energy calculations were performed using the computational resource of the Institute for Computational Biomedicine at Weill Cornell Medical College. In the initial stages of this work, M.V.L. was supported by the National Institutes of Health under Ruth L. Kirschstein National Research Service Award F31DA035533.

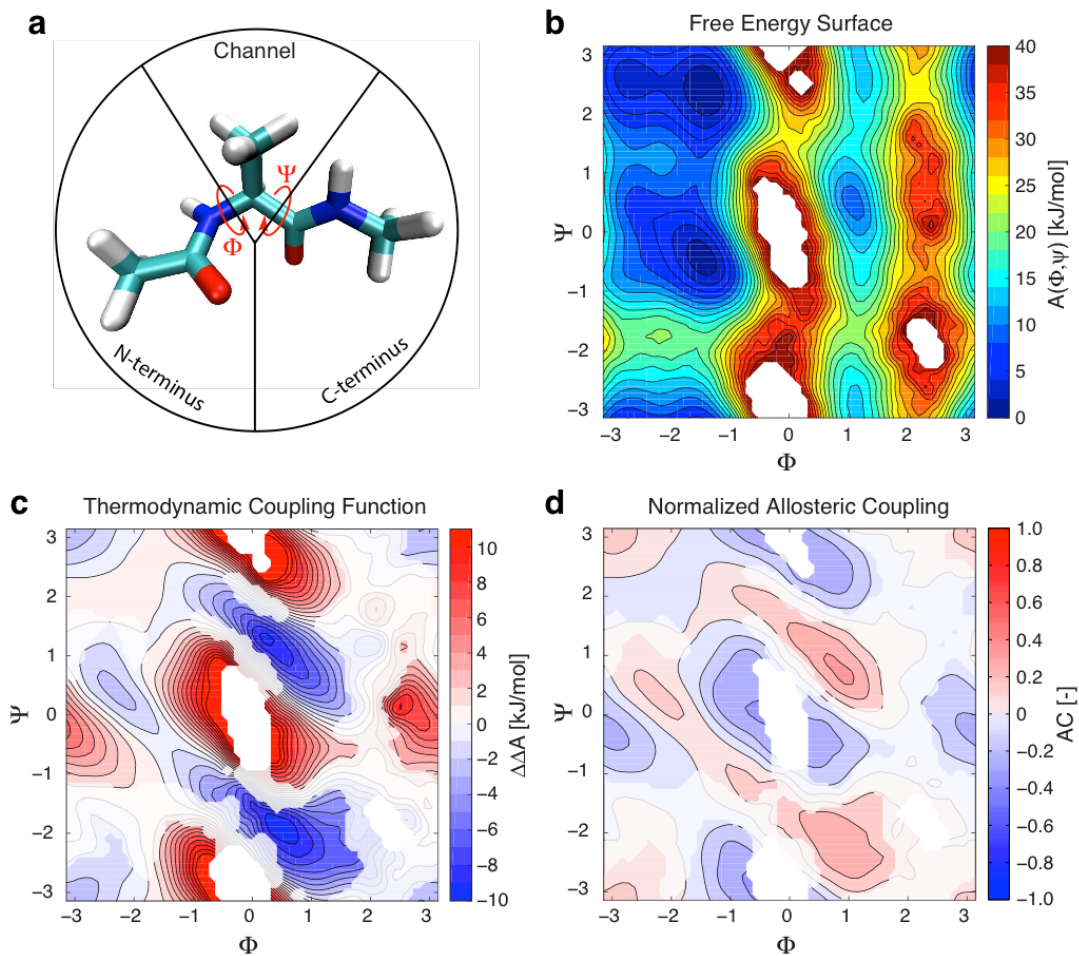

**Figure 1:** Allostery in the alanine dipeptide. (a) The alanine dipeptide molecule with the backbone dihedral angles $\Phi$ and $\Psi$ indicated by arrows. The molecule is partitioned in three domains as indicated by the black lines (see text for details). (b) Free energy surface $A(\Phi,\Psi)$ calculated according to Eq. (13). In all panels, $\Phi$ and $\Psi$ are expressed in radians. (c) The allostery landscape representing the thermodynamic coupling between CVs $\Phi$ and $\Psi$, calculated according to Eq. (16). (d) The normalized allosteric coupling, calculated according to Eq. (19). In panels (c)

and (d), greyed-out regions represent data that are not surely different from zero, based on its 95%-confidence interval estimated by bootstrapping (see Methods).

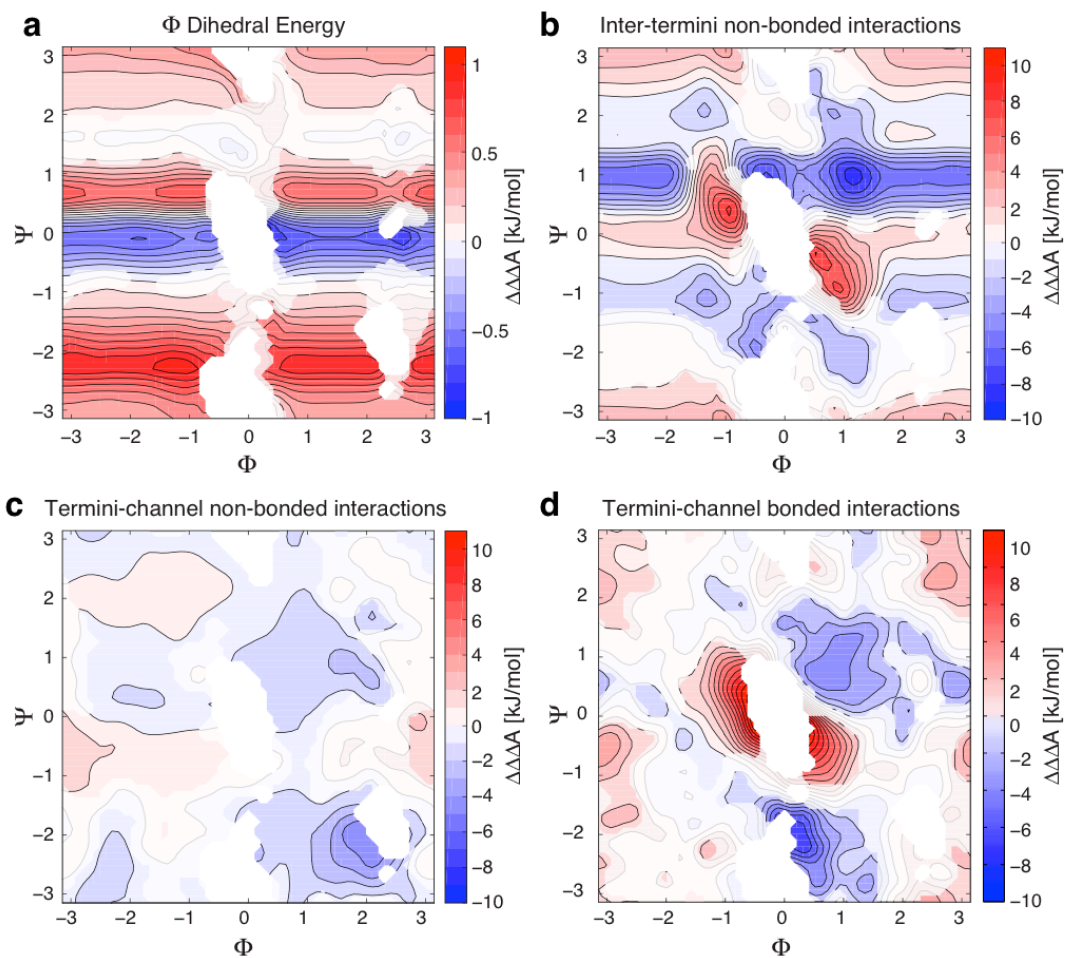

**Figure 2**: Contributions of specific interactions to the allosteric coupling function of the alanine dipeptide, calculated according to Eq. (30). (a) Contribution of the dihedral energy term corresponding to the definition of the angle $\Phi$ (C-N-C$_\alpha$-C). The corresponding interaction energy as a function of $\Phi$ and $\Psi$, unbiased free energy surface, the perturbed allosteric coupling function and the associated AC are shown in Fig. S2. (b) Contribution of the non-bonded interaction energy between the termini. Additional plots in Fig. S3. (c) Contribution of the non-bonded interactions between the termini and the channel. Additional plots in Fig. S4. (d) Contribution of the bonded interactions involving atoms from both the termini and the channel. Additional plots in Fig. S5. In all panels, greyed-out regions represent data that are not surely different from zero, based on its 95%-confidence interval estimated by bootstrapping (see Methods).

## SUPPLEMENTARY INFORMATION

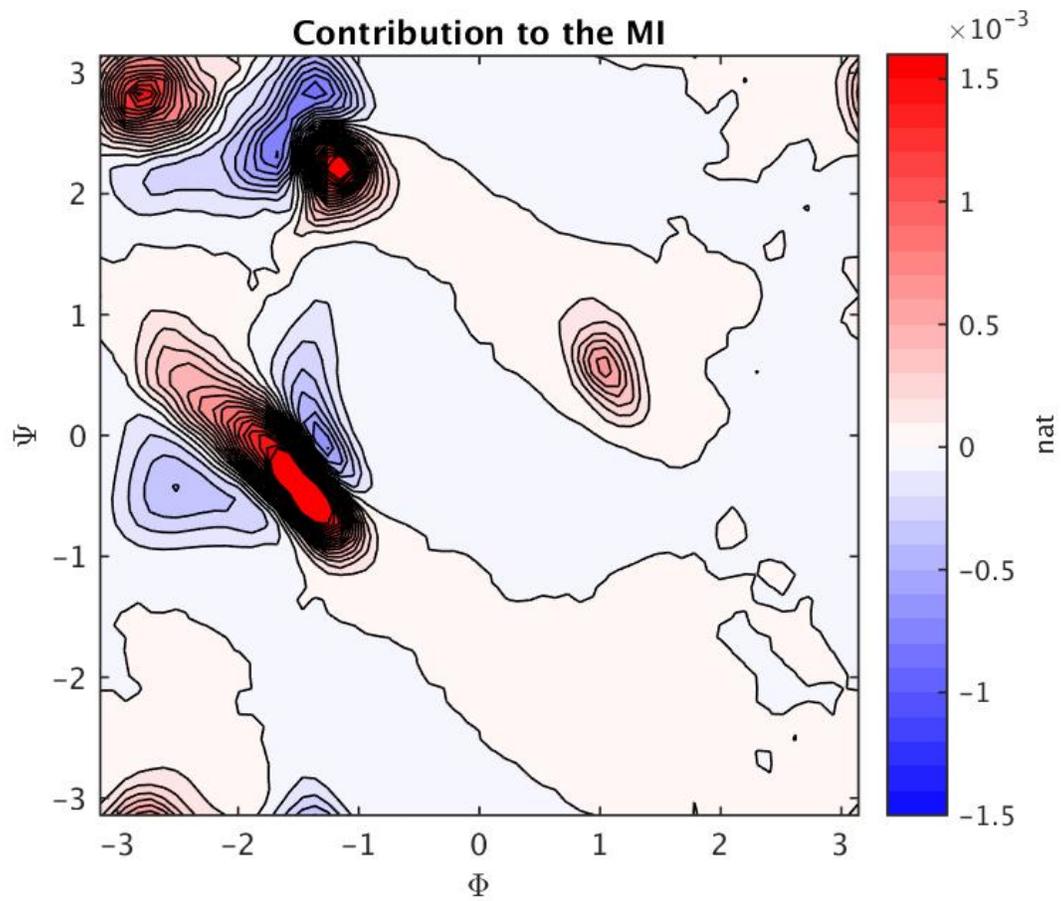

**Figure S1:** The contribution of each bin of $(\Phi,\Psi)$ space to the mutual information. Units are shown in nats.

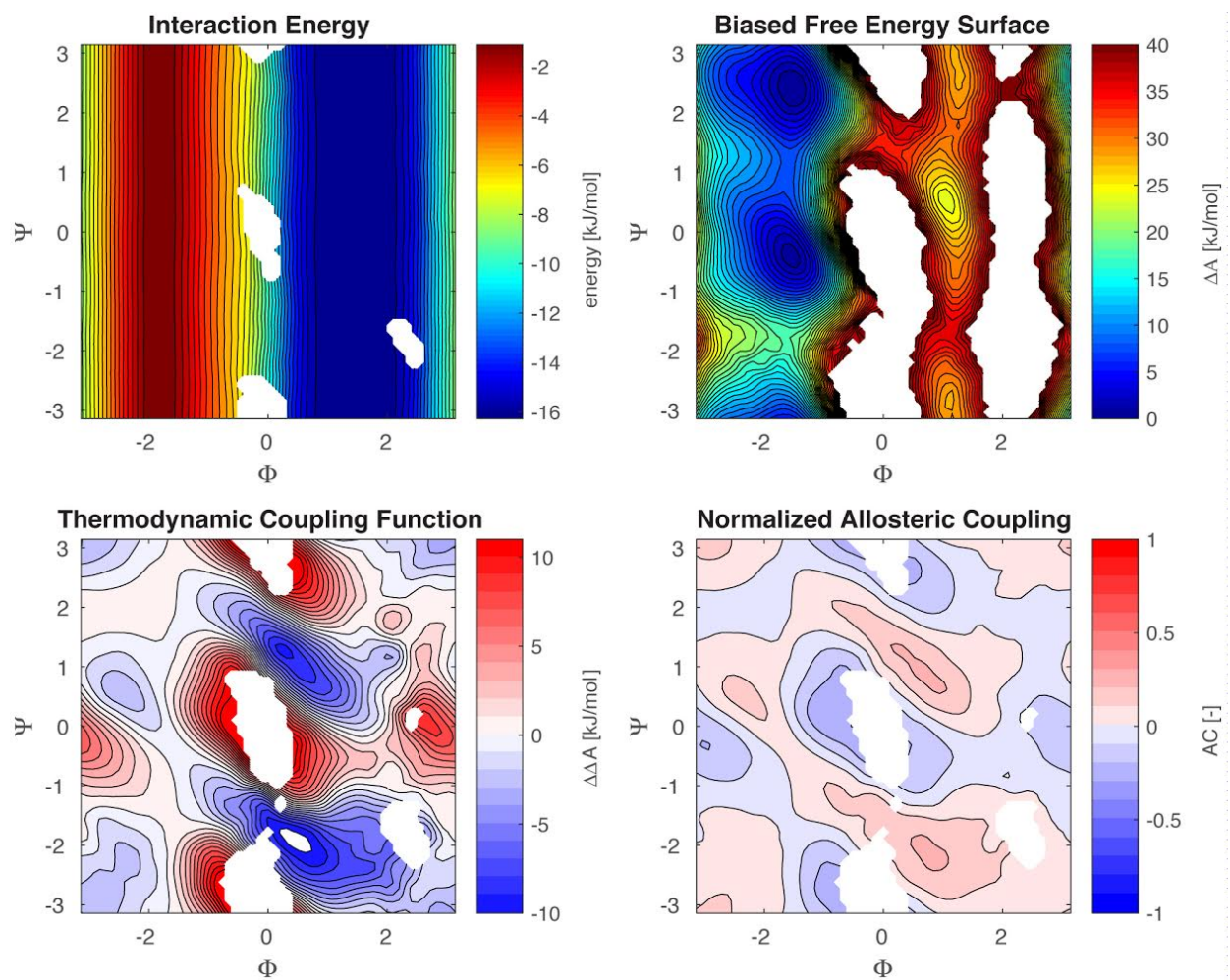

**Figure S2:** Perturbation approach to characterize the contribution of the $\Phi$ dihedral energy term to the allosteric coupling function of the alanine dipeptide. (a) Average interaction energy in bins of the $(\Phi, \Psi)$ space. (b) Unbiased free energy surface according to Eq. (28). (c) Perturbed allosteric coupling function, according to Eq. (31). (d) AC corresponding to the perturbed allosteric function.

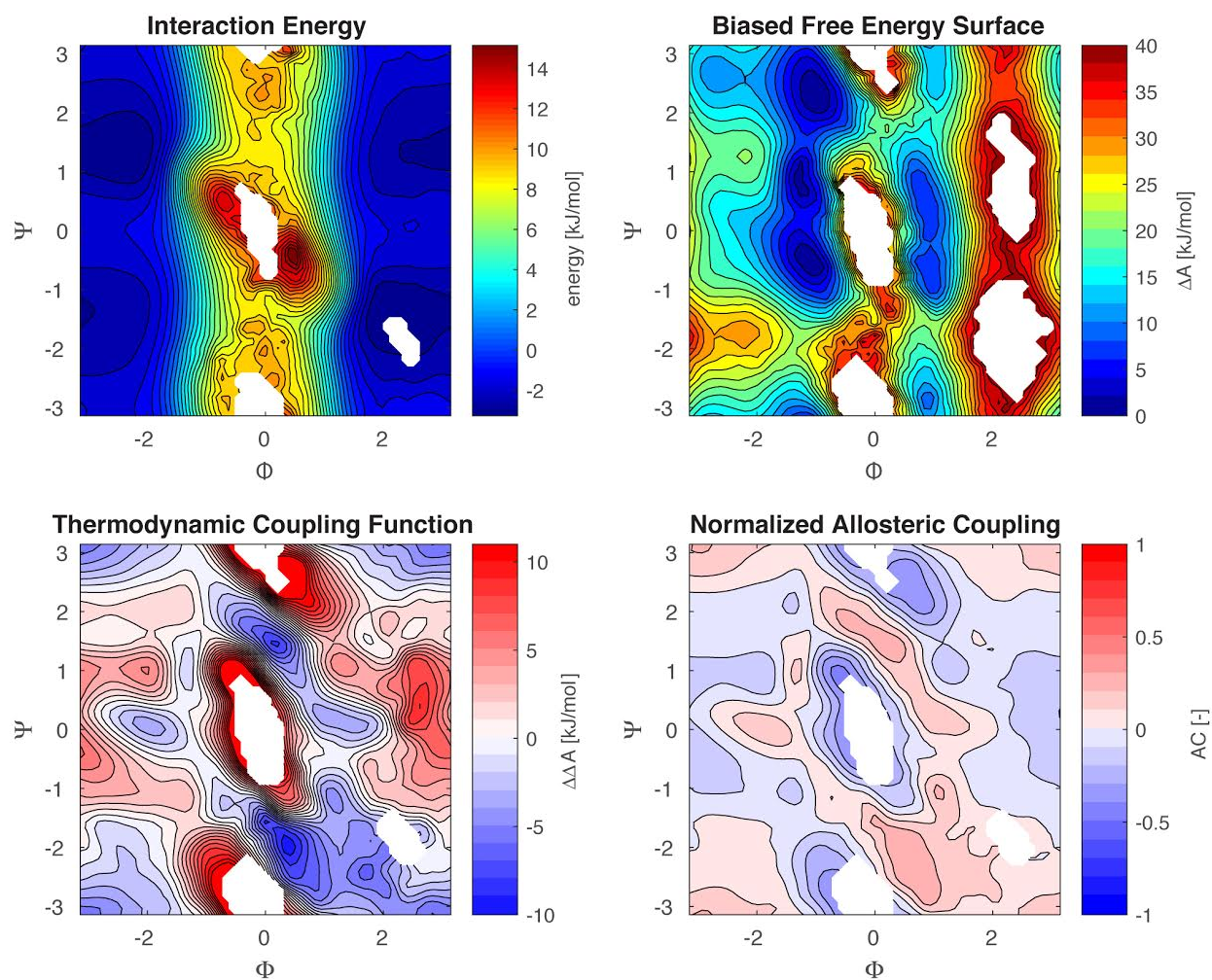

**Figure S3**: Perturbation approach to characterize the contribution of the direct non-bonded interaction energy terms between the N-terminus and the C-terminus of the alanine dipeptide. (a) Average interaction energy in bins of the $(\Phi, \Psi)$ space. (b) Unbiased free energy surface according to Eq. (28). (c) Perturbed allosteric coupling function, according to Eq. (30). (d) AC corresponding to the perturbed allosteric function.

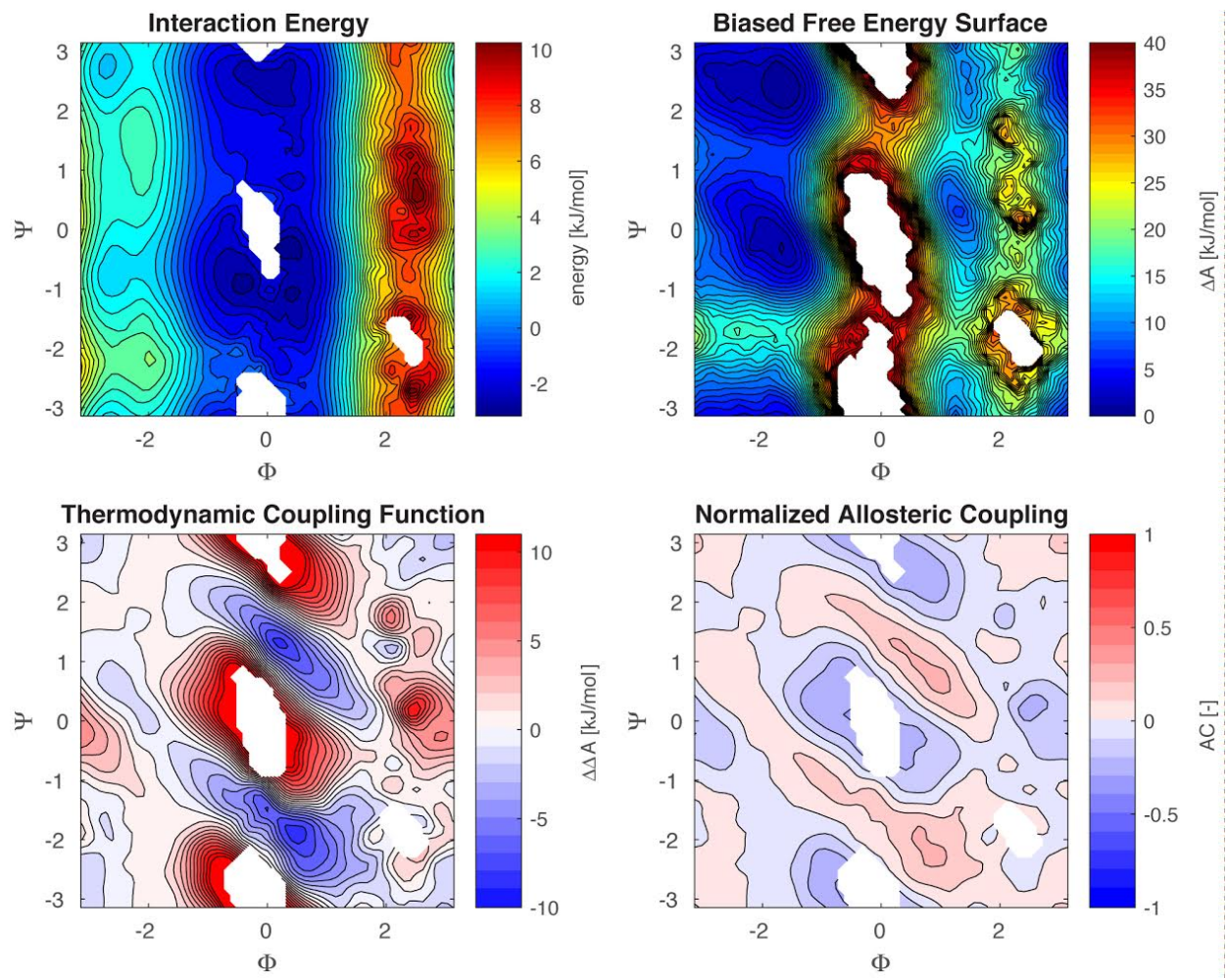

**Figure S4**: Perturbation approach to characterize the contribution of the non-bonded interaction energy terms between the N-terminus and the channel, and between the C-terminus and the channel of the alanine dipeptide. (a) Average interaction energy in bins of the $(\Phi, \Psi)$ space. (b) Unbiased free energy surface according to Eq. (28). (c) Perturbed allosteric coupling function, according to Eq. (30). (d) AC corresponding to the perturbed allosteric function.

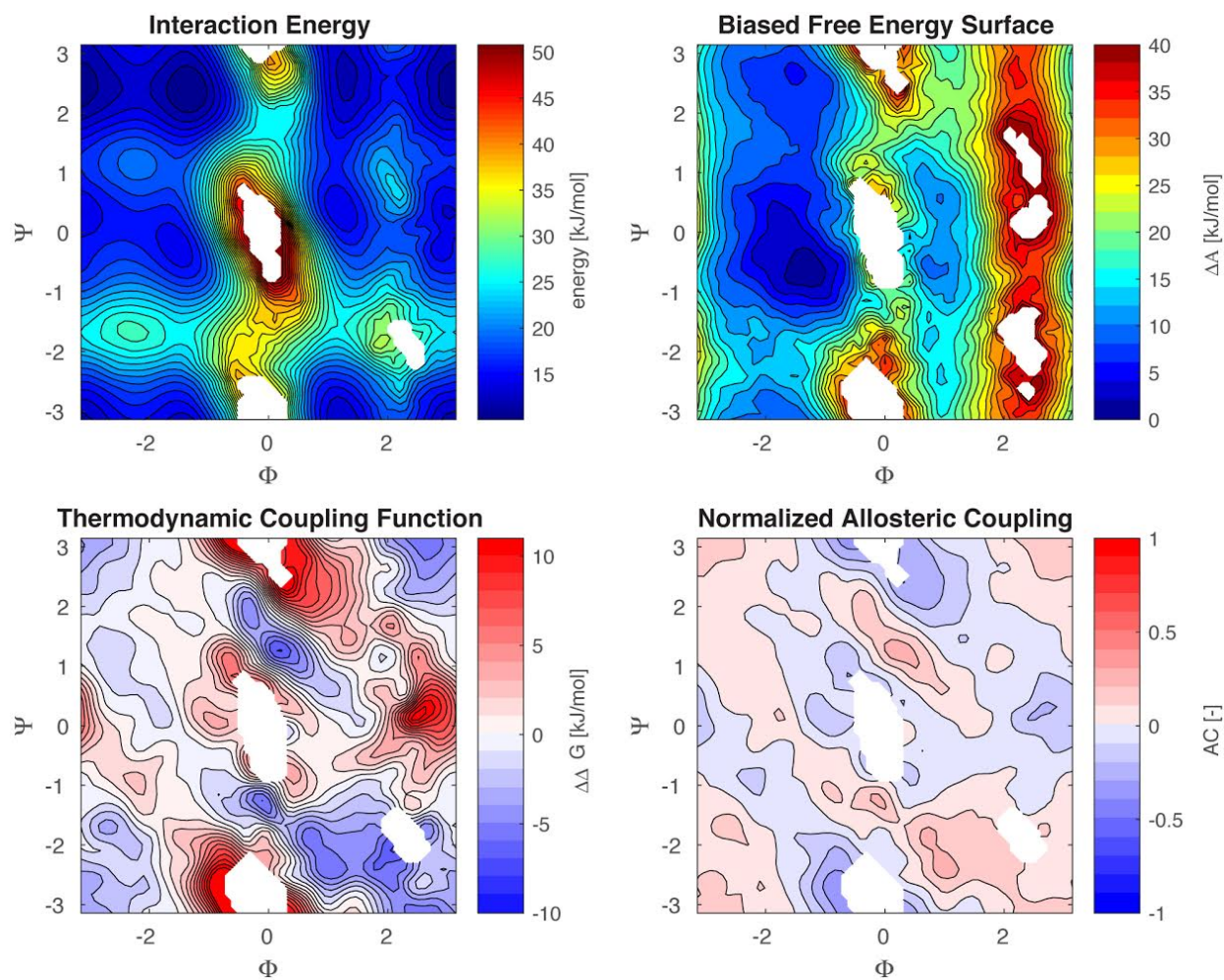

**Figure S5**: Perturbation approach to characterize the contribution of the bonded interaction energy terms involving atoms of the channel and of either the N-terminus or the C-terminus of the alanine dipeptide. (a) Average interaction energy in bins of the $(\Phi,\Psi)$ space. (b) Unbiased free energy surface according to Eq. (28). (c) Perturbed allosteric coupling function, according to Eq. (30). (d) AC corresponding to the perturbed allosteric function.